\documentstyle[12pt,aaspp4,psfig]{article}
\def\ltsima{$\; \buildrel < \over \sim \;$}
\def\simlt{\lower.5ex\hbox{\ltsima}}
\def\gtsima{$\; \buildrel > \over \sim \;$}
\def\simgt{\lower.5ex\hbox{\gtsima}}
\def\kms{{\rm\,km\,s^{-1}}}
\def\kpc{{\rm\,kpc}}

\def\msun{{\rm\,M_\odot}}

\def\s{\ifmmode \widetilde \else \~\fi}
\def\={\overline}

\def\spose#1{\hbox to 0pt{#1\hss}}

\def\etal{{\it et al.\ }}

\def\lta{\mathrel{\spose{\lower 3pt\hbox{$\mathchar"218$}}
     \raise 2.0pt\hbox{$\mathchar"13C$}}}
\def\gta{\mathrel{\spose{\lower 3pt\hbox{$\mathchar"218$}}
     \raise 2.0pt\hbox{$\mathchar"13E$}}}
\def\Dt{\spose{\raise 1.5ex\hbox{\hskip3pt$\mathchar"201$}}}	% upper case
\def\dt{\spose{\raise 1.0ex\hbox{\hskip2pt$\mathchar"201$}}}	% lower case

\def\=={\equiv}

\def\dotsfill{\leaders\hbox to 1em{\hss.\hss}\hfill}

\slugcomment{Accepted by The Astrophysical Journal Letters}

\lefthead{Ibata \etal}
\righthead{Tidal stream in front of the LMC?}

\begin{document}

\title{Re-examination of the possible tidal stream in front of the LMC}

% Notice that each of these authors has alternate affiliations, which
% are identified by the \altaffilmark after each name.  The actual alternate
% affiliation information is typeset in footnotes at the bottom of the
% first page, and the text itself is specified in \altaffiltext commands.
% There is a separate \altaffiltext for each alternate affiliation
% indicated above.

% The abstract environment prints out the receipt and acceptance dates
% if they are relevant for the journal style.  For the aasms style, they
% will print out as horizontal rules for the editorial staff to type
% on, so long as the author does not include \received and \accepted
% commands.  This should not be done, since \received and \accepted dates
% are not known to the author.

\author{
Rodrigo A. Ibata\altaffilmark{1}, 
Geraint F. Lewis\altaffilmark{2} \&
Jean-Philippe Beaulieu\altaffilmark{3}}

\altaffiltext{1}
{European Southern Observatory 
Karl Schwarzschild Stra\ss e 2, D-85748 Garching bei M\"unchen, Germany \nl
Electronic mail: ribata@eso.org}

\altaffiltext{2}{ 
Fellow of the Pacific Institute for Mathematical Sciences 1998-1999, \nl
Dept. of Physics and Astronomy, University of Victoria, Victoria, B.C., Canada 
\& \nl 
Astronomy Dept., University of Washington, Seattle WA, U.S.A.
\nl
Electronic mail: {\tt gfl@uvastro.phys.uvic.ca} \nl 
Electronic mail: {\tt gfl@astro.washington.edu}}

\altaffiltext{3}
{Kapteyn Astronomical Institute 9700 AV Groningen, The Netherlands \nl
Electronic mail: beaulieu@astro.rug.nl}

\begin{abstract}
It has recently been suggested that the stars in a vertical extension of the
red clump feature seen  in LMC color-magnitude   diagrams could belong to  a
tidal stream of material located in front of that  galaxy.  If this claim is
correct,    this   foreground concentration    of    stars  could contribute
significantly  to the rate of gravitational  microlensing events observed in
the   LMC   microlensing experiments.   Here     we present radial  velocity
measurements of stars  in   this  so-called ``vertical red    clump''  (VRC)
population.    The observed stellar  sample, it  transpires, has typical LMC
kinematics.  It is shown  that it is   improbable that an intervening  tidal
stream should have  the same distribution of  radial velocities as the  LMC,
which is consistent with  an earlier study  that showed that the VRC feature
is  more  likely a  young  stellar  population  in   the main  body  of that
galaxy. However,   the kinematic   data  do  not  discriminate   against the
possibility that the VRC is an LMC halo population.
\end{abstract}

% The different journals have different requirements for keywords.  The
% keywords.apj file, found on aas.org in the pubs/aastex-misc directory, 
% contains a list of keywords used with the ApJ and Letters.  These are 
% usually assigned by the editor, but authors may include them in their 
% manuscripts if they wish. 


% That's it for the front matter.  On to the main body of the paper.
% We'll only put in tutorial remarks at the beginning of each section
% so you can see entire sections together.

% In the first two sections, you should notice the use of the LaTeX \cite
% command to identify citations.  The citations are tied to the
% reference list via symbolic KEYs.  We have chosen the first three
% characters of the first author's name plus the last two numeral of the
% year of publication.  The corresponding reference has a \bibitem
% command in the reference list below.
%
% Please see the AASTeX manual for a more complete discussion on how to make
% \cite-\bibitem work for you.   

\section{Introduction}

Approximately 16  microlensing    events  have  been  found by    the  MACHO
collaboration  (e.g.   \markcite{alcock97a}Alcock \etal\  1997a) in  their 4
year  study of  microlensing  towards the  LMC (\markcite{cook98}Cook 1998).
Since this  event rate  exceeds the microlensing   event rate expected  from
standard models of  the Galactic disk, thick  disk  and spheroid components,
they conclude that a  significant fraction --- most  likely half --- of  the
mass  in the Milky Way  is to be found in  massive  baryonic compact objects
which are distributed in  an extended halo around  the Galaxy.  The combined
EROS and MACHO  databases rule out the  possibility that  the lensing masses
are sub-stellar (in the range $3.5  \times 10^{-7} M_\odot <  M < 4.5 \times
10^{-5}  M_\odot$,  \markcite{alcock98}Alcock \etal\   1998); and  the  most
probable mass appears     to   be  in  the     regime   typical of     stars
(\markcite{alcock97a}Alcock \etal\ 1997a).

However,  several studies have proposed alternatives  to the conclusion that
the lensing  masses are genuine halo   objects.  For instance,  some of them
could   belong    to normal  stellar     populations   in  the   LMC  itself
(\markcite{sahu94}Sahu   1994, \markcite{alcock97b}Alcock \etal\     1997b),
though  it  appears  that  this  possibility alone  cannot  account  for the
observed event rate.  Alternatively, either  the lenses or the sources could
belong to   a tidal  structure  either in   front  of,  or behind,   the LMC
(\markcite{zhao98a}Zhao  1998a); such a  structure    could be a   disrupted
remnant of a former dwarf satellite galaxy or a tidally stripped part of the
LMC  itself.   It  is  possible that  the  combination  of   such ``normal''
microlensing events could account for the total observed event rate.  If so,
the case that baryonic dark matter has been detected in the LMC microlensing
experiments would  be  substantially  weakened.  Observational evidence  for
these objections was presented in 1997,  when \markcite{zl97}Zaritsky \& Lin
suggested that  a vertical   feature  in  the color-magnitude  diagram  seen
directly above  the LMC   ``red clump''   could  be due  to an   intervening
population about $20 \kpc$ in front of the LMC.  The color-magnitude diagram
of a central LMC field is shown in Figure~1; we draw attention to the region
in color-magnitude  parameter space of  the  Zaritsky \&  Lin ``vertical red
clump'' (VRC) feature by surrounding it with a box.  The boxed region covers
the color-magnitude range ${\rm 0.49 < V-R < 0.6}$, $17.85  < V < 18.7$; the
color  limits were chosen   visually to encompass  the  feature of interest,
while   the magnitude limits  help us  select  possible foreground red clump
stars  between $\sim    10\kpc$ to  $\sim   25\kpc$  in front of   the  LMC.
\markcite{beaulieu98}Beaulieu  \& Sackett (1998) confirmed   photometrically
the existence of the VRC,  but showed that the  CMD-structure of the feature
is consistent with being  a young stellar population  within the LMC itself.
The observed   characteristics   and densities  in   the VRC  are  precisely
reproduced  with  a  simple model    and a  few  basic assumptions:  stellar
evolution  models  from  \markcite{ber92}Bertelli    \etal\ (1994), an   LMC
distance  modulus of 18.3~mag, an  IMF with a slope  of -2.8  and a constant
star formation history over the last gigayear.

We set out to perform  a complementary investigation of  the nature of these
VRC stars, using radial velocity measurements to determine the kinematics of
the population, as this provides  a direct test  of the competing hypotheses
that does not rely  on any  modeling or  stellar evolution  calculation.  In
particular,  any  discrepancy  in  the  kinematic  properties    of the  VRC
population with respect to the LMC would provide a clear indication that the
line  of sight  to  the LMC   contains some  additional  population that  is
distinct from the LMC.

\section{Spectroscopic observations}

On May 3rd,  1998, we used the  ARGUS bench-mounted multi-fiber spectrograph
at the 4m  Cerro Tololo Interamerican Observatory (CTIO)  to observe a field
in  the central    regions of the  LMC.   The   KPGLD3 grating  was  used in
conjunction  with the   LORAL3k CCD detector    to give  spectra with  ${\rm
1.1\AA/{pixel}}$ resolution,   covering  the range   between  $4000\AA$  and
$7500\AA$.

With ARGUS,  24 fibers can be placed  over a circular field of approximately
40' in diameter,  though the fiber  positioners are not  allowed to touch or
cross over. Each positioner also has a secondary fiber with which allows the
sky spectrum to be  probed close to the target  of interest.  Unfortunately,
the region where our CCD photometry had accurately calibrated astrometry was
quite  small  (approximately $6'\times 6'$),  which  meant that   it was not
possible to place all  fibers on objects  of interest.  We exposed two fiber
setups, observing 24 stars that lie within the boxed region of Figures~1 and
2, plus 16 stars that lie outside that region.  The  objects for which ARGUS
spectra were obtained  are marked with  a  large dot in  the color-magnitude
diagram in Figure~2.   Each  fiber setup was  exposed for  3600~sec,  giving
S/N$\sim 10$ for the faintest stars observed.  The sky spectrum, constructed
by performing a  median of the sky-fiber  spectra,  was subtracted from  the
(fiber-response corrected) stellar spectra. Judging by the size of remaining
emission  lines  in the  spectra, the sky   subtraction is accurate to $\sim
5$\%.  Details of the data-reduction  techniques used with this instrumental
setup can be found in \markcite{sun93}Suntzeff \etal\ (1993).

The  resulting velocity data are listed  in Table~1, which contains accurate
coordinates,   photometry,   Tonry-Davis   cross-correlation   $R$    values
(\markcite{tonry79}Tonry \&  Davis 1979), heliocentric radial velocities and
radial   velocity   uncertainties for   40   stars.    The  radial  velocity
uncertainties in Table~1 are derived from  the cross-correlation peak fit of
the survey stars with  the template (the G8 \rm{III}  star HR6468  --- which
gave the best cross-correlation  $R$ values of  all the observed templates),
and are therefore only indicative.

In  the top panel of Figure~3,  we show the  velocity distribution of the 24
spectroscopically observed   stars  that  lie inside   the boxed  region  of
Figures~1  and 2.  The    maximum-likelihood  Gaussian fit to the   velocity
distribution of this sample is found to have a mean of $267 \pm 5 \kms$, and
a velocity dispersion of  $25 \pm  4 \kms$.  The  bottom panel  displays the
velocity distribution of the 16 remaining  stars.  The Gaussian function fit
to these stars has a mean of $270 \pm 4 \kms$ and  a dispersion of $19 \pm 4
\kms$.   The uncertainties on  these    quantities have been estimated    by
boot-strap resampling of the  datasets.  Note that the  velocity dispersions
derived  above are   dominated  by  the large  instrumental   uncertainty in
individual velocity measurements.   To determine   an  upper limit on    the
velocity  dispersions   of    the two  populations,   we   performed further
maximum-likelihood fits  by including a  Gaussian noise model for each datum
(with $\sigma$ equal to   the   cross-correlation uncertainties listed    in
Table~1). The resulting upper limit, at the 90\%  confidence level, is found
to be  $17 \kms$, while  that of the non-VRC  population is $18  \kms$ (note
that these  values depend critically on  whether  the velocity uncertainties
have been well estimated or not).

\section{Discussion and summary}

The heliocentric radial velocity of the \ion{H}{1} at  the center of gravity
of the LMC has  been found to be $274  \pm 6 \kms$ (\markcite{luks92}Luks \&
Rohlfs 1992), while the velocity dispersion of  the carbon stars that reside
in  the LMC disk is  $\sim  15 \kms$ (\markcite{kunkel97}Kunkel \etal\ 1997)
independent of position. Evidently, the kinematic parameters of both our VRC
and non-VRC   samples are  very  similar to   these values,  supporting  the
hypothesis  that the VRC  is a stellar population  of  the LMC. However, the
small velocity dispersion of the VRC population is  also consistent with the
expected velocity dispersion of Galactic  tidal streams, judging from N-body
disruption simulations of dwarf galaxies  (see e.g.  \markcite{ibata98}Ibata
\& Lewis 1998).  Below, we use the small difference  between the mean radial
velocity of the LMC and  the VRC populations  to examine these two  possible
pictures.

The first possibility is that the VRC is a tidal stream that is unrelated to
the LMC.    We can obtain  a  rough  estimate of  how unlikely  the observed
coincidence of mean radial velocities is  in this scenario, by assuming that
the hypothesised tidal stream is a Galactic halo tracer (i.e.  a random halo
particle); if so,  it  must be drawn  randomly  from  the halo  distribution
function.  A reasonable model for  the distribution function of the Galactic
halo  has been   presented  by  \markcite{evans94}Evans   (1994).  For   the
parameters of this ``power law halo'' model, we adopt those corresponding to
the  spherical Galactic halo  model given  in \markcite{ej94}Evans \& Jijina
(1994).  With these assumptions, and taking the mean velocity of our non-VRC
sample as indicative of the mean velocity of  the LMC in the observed field,
we find  that  the probability that the  difference  in mean radial velocity
between the LMC and the tidal stream should be  as low as $3  \pm 6 \kms$ is
very small, $\sim 1$\%.   Note however, that the  use of the Evans \& Jijina
model in deriving this estimate probably exaggerates the unlikelihood of the
observed velocity coincidence; most outer Galactic halo objects are known to
be clustered on a great  circle (see e.g.  \markcite{lynden95}Lynden-Bell \&
Lynden-Bell 1995),  so the two radial  velocities in  the above scenario are
likely to be correlated.

Alternatively, the  VRC could be  a  population of stars that  is physically
associated to the LMC. For instance, it is easy to imagine that some earlier
galactic interaction (e.g.  \markcite{zhao98b}Zhao 1998b) could have ejected
stars from  the main body of the   LMC, though without  sufficient energy to
unbind  them.  A further possibility  is  that the VRC   are LMC halo stars,
scattered  between $\sim 10\kpc$  to $\sim 25\kpc$ in  front of that galaxy.
LMC halo  objects should have a  velocity dispersion approximately  equal to
the virial velocity dispersion  of the system.   Given the LMC mass of $\sim
1.5     \times  10^9   \msun$     and    half-mass  radius    $\sim   7\kpc$
(\markcite{kunkel97}Kunkel \etal\ 1997),  the virial velocity dispersion  is
$\sim 60 \kms$, and the line-of-sight velocity dispersion is correspondingly
a factor of $1/\sqrt{3}$ smaller, $\sim 35 \kms$.  Thus it would appear that
the coincidence  between the systemic velocity of  the  LMC and that  of the
hypothesised stream should be unlikely; or alternatively, if the VRC is part
of a diffuse halo population,  it would appear  that the velocity dispersion
upper  limit of $17  \kms$  is  inconsistent  with the expected  dispersion.
However, there  will be a kinematic bias  in our sample  if  either of these
scenarios   is correct.  This effect is   particularly pronounced due to the
location of the ARGUS field on the sky, close to the dynamical center of the
LMC.    For instance, if  the  distribution function of  the  LMC halo has a
substantial fraction of radial orbits, the  fact that we have selected stars
at large distances  in front of  the LMC  means  that we are also  selecting
stars close to turnaround,  where  they will preferentially  have velocities
close to the  systemic velocity of  their  parent population.   So the small
velocity  dispersion of  the VRC and  the  small difference between the mean
velocities  of the VRC  and the LMC  cannot  be used to discriminate against
these models.

We have presented new kinematic data on a population of  stars that has been
seen   in photometric   studies    of the center     of the  LMC.  In    the
color-magnitude   diagram, these stars   extend  vertically upwards from the
well-known red clump.   Being a rare phenomenon, \markcite{zl97}Zaritsky and
Lin (1997) hypothesised that this population belongs to a previously unknown
stellar stream  located $\sim  20  \kpc$ in front  of  the LMC.  The stellar
components of the stream  are extremely interesting  due to their ability to
bias LMC microlensing experiments.  Strong doubts have  already been cast on
the arguments used  by \markcite{zl97}Zaritsky  and  Lin (1997) to reject  a
stellar       evolution        interpretation       to    this       feature
(\markcite{beaulieu98}Beaulieu  \&  Sackett   1998),  since  simple  stellar
population modeling  can   reproduce from 75\%   to  100\% of   the observed
densities.  The radial velocity data   presented in this contribution  shows
unambiguously   that the ``vertical   red  clump'' population  has identical
kinematic properties to the LMC itself.  If the VRC were a tidal stream, not
associated  to  the LMC, the probability  of  this chance velocity alignment
appears  to be quite small.   However, our data  do not discriminate against
the possibility that  the VRC is a  stellar population gravitationally bound
to the LMC.  Whether  this or another  stellar  population, perhaps a  tidal
streamer very close to the LMC  (\markcite{zhao98a}Zhao 1998a), is affecting
the observed microlensing rate, still remains an open question.

\acknowledgments

This work was based on observations taken at the Cerro Tololo Inter-American
Observatory, which is operated by AURA Inc.\ under  contract to the National
Science Foundation. We thank the anonymous  referee for the pointing out the
kinematic bias of our sample.

\begin{deluxetable}{ccccccccc}
\footnotesize
\tablenum{1}
\tablecaption{CTIO Radial Velocity Data}
\tablewidth{0pt}
\tablehead{
\colhead{RA} & \colhead{DEC} & \colhead{V} & \colhead{${\rm V - R}$} & 
\colhead{$R_{\rm TD}$} & \colhead{$V_{helio}$} & \colhead{$\Delta V_{helio}$} & VRC \nl
\colhead{(2000)} & \colhead{(2000)} & \colhead{(mag)} & \colhead{(mag)} & 
 & \colhead{(km/s)} & \colhead{(km/s)} & member? }
\startdata
5  15 13.9 & $-$68  45 ~4.24 &  17.14 &  0.21 & 17.68 &  234. &  ~9. & n \nl
5  15 ~4.3 & $-$68  45 11.63 &  17.47 &  0.55 & 11.75 &  270. &  14. & n \nl
5  14 31.6 & $-$68  45 15.08 &  18.07 &  0.42 & ~5.88 &  273. &  27. & n \nl
5  14 53.5 & $-$68  45 18.40 &  17.43 &  0.46 & 16.01 &  279. &  10. & n \nl
5  14 45.5 & $-$68  45 20.07 &  17.72 &  0.37 & ~8.39 &  253. &  18. & n \nl
5  14 59.6 & $-$68  45 22.24 &  17.89 &  0.52 & ~8.36 &  233. &  19. & y \nl
5  14 17.4 & $-$68  45 33.64 &  18.48 &  0.59 & 16.33 &  251. &  10. & y \nl
5  14 36.3 & $-$68  45 47.10 &  17.31 &  0.53 & 13.17 &  286. &  13. & n \nl
5  14 14.5 & $-$68  46 ~3.30 &  17.14 &  0.42 & 10.97 &  263. &  15. & n \nl
5  14 23.2 & $-$68  46 12.70 &  18.38 &  0.54 & ~6.70 &  297. &  21. & y \nl
5  15 ~0.3 & $-$68  46 17.27 &  18.04 &  0.57 & 14.41 &  255. &  12. & y \nl
5  14 49.9 & $-$68  46 31.21 &  17.95 &  0.54 & 10.46 &  249. &  17. & y \nl
5  14 33.6 & $-$68  46 39.44 &  18.01 &  0.55 & 10.49 &  256. &  16. & y \nl
5  15 ~8.5 & $-$68  46 42.50 &  17.91 &  0.55 & 14.48 &  273. &  11. & y \nl
5  14 21.7 & $-$68  46 50.10 &  18.47 &  0.60 & ~6.65 &  266. &  23. & y \nl
5  14 15.7 & $-$68  47 ~2.75 &  18.06 &  0.52 & ~9.31 &  258. &  17. & y \nl
5  14 54.6 & $-$68  47 10.47 &  18.23 &  0.57 & ~7.85 &  253. &  20. & y \nl
5  14 22.5 & $-$68  47 28.43 &  18.04 &  0.54 & ~9.99 &  301. &  16. & y \nl
5  14 40.2 & $-$68  47 35.13 &  17.18 &  0.51 & 21.43 &  267. &  ~8. & n \nl
5  15 ~8.9 & $-$68  47 40.57 &  17.96 &  0.53 & ~9.15 &  227. &  18. & y \nl
5  15 ~7.2 & $-$68  47 41.18 &  18.16 &  0.54 & ~6.03 &  250. &  23. & y \nl
5  14 21.7 & $-$68  48 ~9.13 &  18.26 &  0.53 & 15.60 &  272. &  11. & y \nl
5  14 26.9 & $-$68  48 ~9.55 &  17.98 &  0.26 & ~5.07 &  249. &  41. & n \nl
5  15 12.0 & $-$68  48 11.01 &  18.37 &  0.56 & ~5.07 &  187. &  37. & y \nl
5  14 25.8 & $-$68  48 41.62 &  17.31 &  0.55 & 20.27 &  263. &  ~8. & n \nl
5  15 11.6 & $-$68  48 57.03 &  17.52 &  0.55 & 17.23 &  288. &  ~9. & n \nl
5  15 15.9 & $-$68  48 57.21 &  17.16 &  0.46 & 14.76 &  265. &  11. & n \nl
5  14 16.2 & $-$68  49 ~1.36 &  17.25 &  0.57 & 15.14 &  309. &  11. & n \nl
5  14 11.8 & $-$68  49 16.10 &  17.54 &  0.52 & 14.66 &  273. &  11. & n \nl
5  15 ~0.0 & $-$68  49 30.35 &  18.16 &  0.52 & ~9.64 &  275. &  19. & y \nl
5  14 16.7 & $-$68  49 37.67 &  18.23 &  0.51 & ~6.11 &  280. &  29. & y \nl
5  14 38.9 & $-$68  49 42.08 &  18.47 &  0.59 & ~7.66 &  251. &  20. & y \nl
5  14 24.2 & $-$68  49 46.84 &  18.28 &  0.55 & 10.03 &  287. &  16. & y \nl
5  14 55.8 & $-$68  49 52.14 &  18.01 &  0.55 & ~9.58 &  262. &  17. & y \nl
5  15 12.8 & $-$68  49 55.98 &  18.47 &  0.60 & 25.19 &  260. &  ~6. & y \nl
5  14 14.4 & $-$68  50 37.62 &  17.28 &  0.59 & 14.70 &  301. &  11. & n \nl
5  14 22.2 & $-$68  50 37.71 &  17.50 &  0.53 & 23.26 &  252. &  ~7. & n \nl
5  14 21.3 & $-$68  50 47.95 &  18.38 &  0.56 & ~7.88 &  303. &  25. & y \nl
5  14 30.7 & $-$68  50 57.88 &  18.48 &  0.56 & ~5.86 &  299. &  24. & y \nl
5  15 12.3 & $-$68  50 59.01 &  18.31 &  0.52 & ~6.30 &  271. &  20. & y 
\enddata
\end{deluxetable}

\begin{figure}[htb]
\psfig{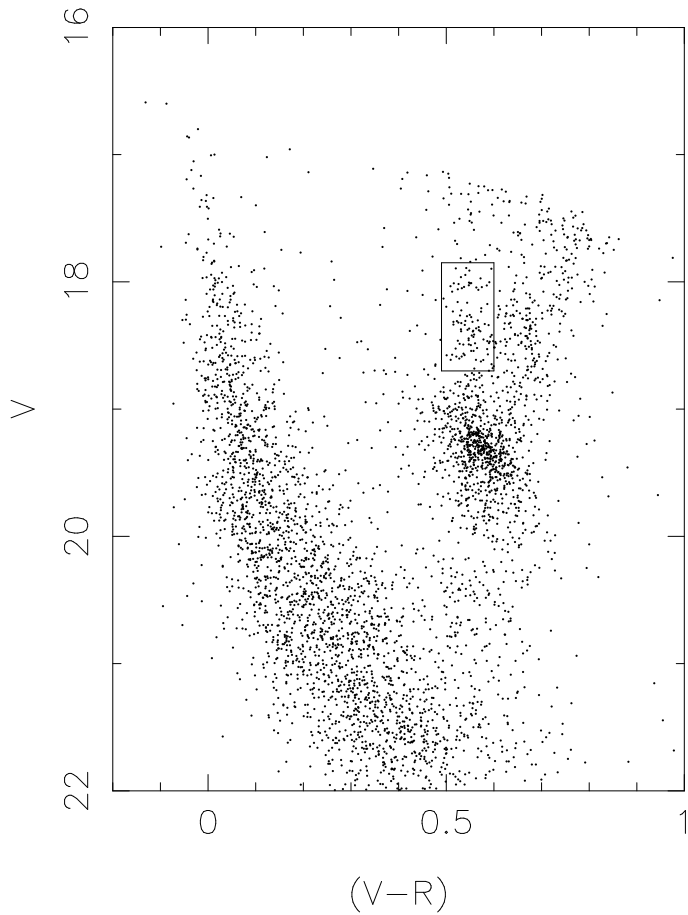}
\caption[  ]{
This    color magnitude   diagram  shows  a   field  at $\alpha=5^h  14'.7$,
$\delta=-68^\circ 49'.9$    (J2000), in the    central  regions of   the LMC
(Beaulieu  \& Sackett 1998). The ``vertical  red clump'' (VRC) identified by
Zaritsky  and Lin (1997)   is clearly seen within   the boxed region in this
diagram  (the limits of  this box were chosen by  ourselves to encompass the
population of interest).}
\end{figure}
\begin{figure}[htb]
\psfig{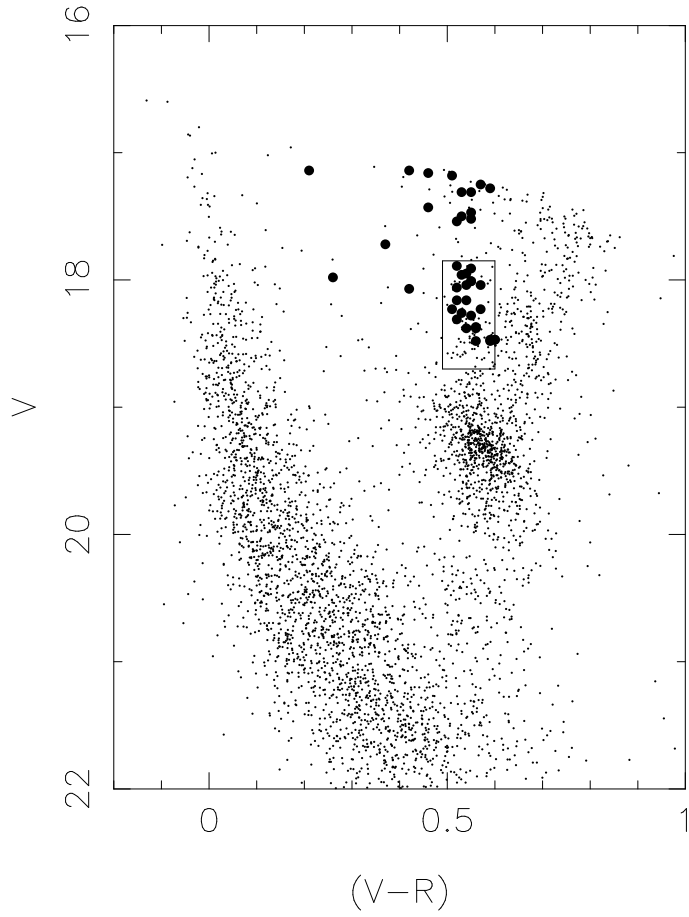}
\caption[  ]{
The color-magnitude positions  of  the spectroscopically observed stars  are
shown superimposed on the  CMD of Figure~1. Of  the 40 stars, 24  lie within
the VRC box.}
\end{figure}

\begin{figure}[htb]
\psfig{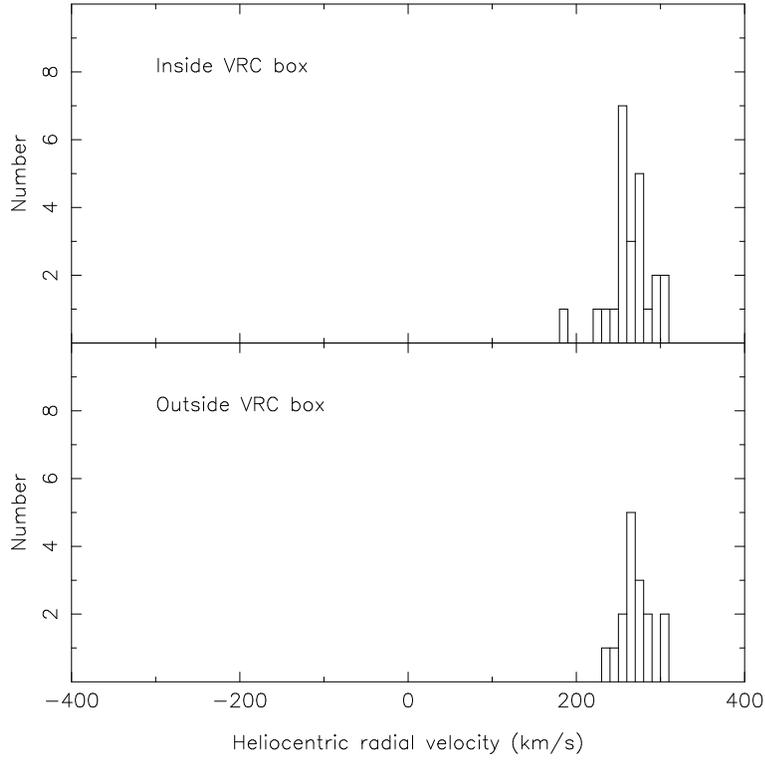}
\caption[   ]{  
The top panel shows the heliocentric radial velocity distribution of the 24
stars that lie in the VRC box in Figure~2.  The velocity distribution of all
other stars is displayed in the bottom panel.}
\end{figure}

\end{document}